\RequirePackage{lineno}
\documentclass[aps,prc,twocolumn, floatfix]{revtex4}
\usepackage{graphicx}
\usepackage{dcolumn}
\usepackage{bm}

\begin{document}


\title{Phase diagram determination at fivefold nuclear compression}

\author{Gao-Chan Yong$^{1,2}$}

\affiliation{
$^1$Institute of Modern Physics, Chinese Academy of Sciences, Lanzhou 730000, China\\
$^2$School of Nuclear Science and Technology, University of Chinese Academy of Sciences, Beijing 100049, China
}

\begin{abstract}

In the standard model of particle physics, the strong force is characterized by the theory of quantum chromodynamics (QCD). It is commonly understood from QCD properties that hadrons, at sufficiently high temperatures or densities, melt into their constituent quarks, thereby undergoing a deconfinement transition to a new phase of quarks and gluons, often referred to as quark matter or quark-gluon plasma (QGP) \cite{qcd00,qcd01}. Although QGP has been observed in relativistic heavy-ion collisions \cite{qgp1,qgp2}, uncertainties remain about when the onset of deconfinement occurs. After comparing simulations from a reliable hadron and quark relativistic transport model with recent data from the STAR experiment, we determined that the onset of the hadron-quark phase transition occurs at about five times nuclear compression, corresponding to temperature $T\sim$ 112 MeV and baryon chemical potential $\mu_{B}\sim$ 586 MeV, in the nuclear matter phase diagram. This discovery has significant implications for the studies of both the early and present universe \cite{ann2006}, including the fraction of dark matter formed in the early universe \cite{bhd2016,bhf1997,pbh20} and the structure and dynamics of neutron stars and their mergers \cite{nature2020}.

\end{abstract}

\maketitle

\begin{figure}[t]
\centering
\includegraphics[width=0.45\textwidth]{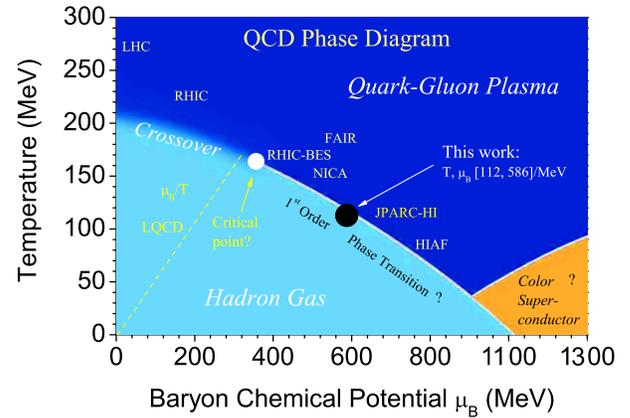}
\caption{A sketch of the QCD phase diagram, which depicts temperature as a function of baryon chemical potential. It includes the conjectured QCD critical point (white point), and the first-order phase transition line (the black point at T $\sim$ 112 MeV and $\mu_{B}$ $\sim$ 586 MeV indicates this work, see text for details). The region predicted by Lattice QCD to follow a smooth crossover is signified by the yellow line up until $\mu_{B}$/T $\leq$ 2. The coverage of several worldwide facilities is roughly labeled around the phase transition boundary line.} \label{ptdia}
\end{figure}

One of the most crucial properties of the theory of quantum chromodynamics (QCD) is asymptotic freedom \cite{jjzy1,jjzy2}, which means that the coupling constant decreases with increasing energy scale. Consequently, it is naturally anticipated that QCD matter at high energy densities undergoes a phase transition from a state with confined hadrons to a new quark-gluon plasma (QGP) state. While lattice QCD has established that the transition at vanishing net-baryon density is a smooth crossover \cite{latt,latt2}, the presence of a first-order transition accompanied by a critical end point (CEP) has been conjectured based on many effective theories \cite{rev1,pt1}, as shown in Figure~\ref{ptdia}. The study of the CEP as well as the first phase transition boundary has become the current focus of numerous research activities worldwide, both theoretically and experimentally \cite{pt1,nara18}. Today, mapping the QCD phase diagram is the major scientific goal of the second phase of the Beam Energy Scan (BES-II) program in heavy-ion collisions at the Relativistic Heavy Ion Collider's STAR Collaboration (RHIC-STAR) \cite{pr2,bes19,besa,besb}. The Compressed Baryonic Matter experiment at the Facility for Antiproton and Ion Research (CBM-FAIR) aims to study the Equation of State (EOS, a mathematical relationship that links thermodynamic quantities such as energy, temperature, pressure, and density) of dense baryonic matter, a possible first-order phase transition, and the existence of the CEP in the baryon-rich domain by measuring rare probes \cite{CBM17,CBM21}. The Nuclotron-based Ion Collider fAcility (NICA) at the Joint Institute for Nuclear Research allows for the study of the EOS of dense baryonic matter and the QCD phase transition by measuring multi-strange hyperons and hypernuclei with the Multi-Purpose Detector (MPD) \cite{NICA19}. Many other facilities, both constructed and under construction worldwide, have related research projects and plans \cite{qm2018}.

The study of the QCD matter phase transition from the Earth to outer space is believed to have vital implications for our understanding of both the early and modern universe \cite{ann2006} in the field of astronomy. Determining whether a hadron-quark phase transition exists in neutron stars (NSs) at central densities several times the nuclear saturation density holds considerable significance for research into neutron star structure \cite{akm1998,nature2020,liapj2020,hu2021,kojo2021,ran2021,xie2021} and gravitational-wave (GW) emission \cite{gw2019,gw2018, gw20182}. The primordial black holes, considered an ideal candidate for cosmic dark matter \cite{bhd2016}, may have formed through pre-existing density fluctuations during the quark-hadron phase transition \cite{bhf1997}.

In Ref.~\cite{mf2022}, the STAR Collaboration posited that the number-of-constituent-quark scaling (NCQS), a model-independent method used to ascertain the presence of quark matter in heavy-ion collisions, is missing. They suggest that a new equation of state (EOS), likely driven by baryonic interactions in regions of high baryon density, is required to mirror the features of the observed negative elliptic flow $v_{2}$, and the positive directed flow $v_{1}$ slope in midcentral Au+Au collisions at a nucleon-nucleon center-of-mass energy of $\sqrt{s_{NN}}$ = 3 GeV. On the other hand, Ref.~\cite{qk2021} observes that pion and proton elliptic flows in Au+Au collisions at $\sqrt{s_{NN}}$ = 4.5 GeV exhibit a trend suggestive of NCQS. These findings indicate that a definitive conclusion may be drawn from the comprehensive statistics of the ongoing BES-II. Consequently, the occurrence of the hadron-quark phase transition in relativistic heavy-ion collisions could potentially be within the $\sqrt{s_{NN}}$ = 3 to 4.5 GeV range. While examining the onset of the hadron-quark phase transition in relativistic heavy-ion collisions, it is crucial to first replicate various hadronic observable measurements using a single transport model, at least within the $\sqrt{s_{NN}}$ = 3 to 4.5 GeV range. Regrettably, there are no current models capable of reproducing all recent STAR experimental observables \cite{qcd2022}. To meaningfully describe experimental data within the BES-II and CBM-FAIR range, transport codes must incorporate physics that allow for the reproduction of all essential measurements \cite{eos2023}. For this reason, in this study, we delve into the $v_{1}$ and $v_{2}$ flows for various hadrons based on the recently updated isospin and momentum-dependent hadron cascade mode in a multi-phase transport model \cite{cas2021,yongrcas2022}. The simulation results are found to adequately replicate various hadronic observable measurements in Au+Au collisions at $\sqrt{s_{NN}}$ = 3 and 4.5 GeV. Furthermore, by analyzing the proton elliptic flow, we ascertain the proportion of quark matter created in Au+Au collisions at $\sqrt{s_{NN}}$ = 4.5 GeV.


In line with the focus of this study, which is the exploration of the emergence of quark matter or, equivalently, the phase-transition boundary of QCD matter from the hadronic phase to the QGP phase, the multi-phase transport (AMPT) model \cite{AMPT2005} has been recently refined. This refinement allows the model to perform pure hadron cascade simulations with hadronic mean-field potentials, in addition to multi-phase transport simulations incorporating both parton and hadron degrees of freedom \cite{cas2021,yongrcas2022}. As a Monte Carlo model which accommodates both parton and hadron transport, the AMPT model comprises four elements: fluctuating initial conditions, partonic interactions, the transition from partonic to hadronic matter, and hadronic interactions \cite{AMPT2005}. This model has been successfully applied to heavy-ion collisions at RHIC and LHC energies \cite{nst2021}. In the employed string melting AMPT model (AMPT-SM), the initial partons are derived through the intermediate decomposition of hadrons created via Lund string fragmentation, following the Heavy Ion Jet INteraction Generator (HIJING) model \cite{wang1,wang2}. The original Lund string fragmentation parameters $a$ = 0.55, $b$ = 0.15/GeV$^{2}$, the strong coupling constant $\alpha_{s}$= 0.33, and the parton cross section $\sigma$ = 3 mb are retained \cite{linab14}. The scatterings of melted partons are represented by the Zhang's Parton Cascade (ZPC) model \cite{zhang1}. After the application of a quark coalescence model for hadronization, subsequent hadronic interactions are depicted by a hadronic cascade based on a relativistic transport (ART) model \cite{art,yongrcas2022}. It should be noted that at lower energies, the currently applied AMPT-SM model incorporates the effects of finite nuclear thickness \cite{thick3, thick1, thick2}.

In order to investigate the potential hadron-quark phase transition in heavy-ion collisions, a pure hadron cascade model (AMPT-HC) has been developed \cite{cas2021,yongrcas2022}. The recently updated AMPT-HC model takes into account the initial density and momentum distributions of nucleons in colliding nuclei, as provided by Skyrme-Hartree-Fock calculations with Skyrme M$^{\ast}$ force parameters \cite{skyrme86}. This model also incorporates recent experimental findings on the nucleon momentum distribution, with a high-momentum tail extending up to approximately 2 times the local Fermi momentum \cite{yongsrc}. An isospin- and momentum-dependent single-nucleon potential (MDI) is utilized, as detailed in Refs.~\cite{yongsrc,spp1,yong20152}.
The employed isospin- and momentum-dependent single-nucleon potential (MDI) can be expressed as follows:
\begin{eqnarray}
U(\rho,\delta,\vec{p},\tau)&=&A_u(x)\frac{\rho_{\tau'}}{\rho_0}+A_l(x)\frac{\rho_{\tau}}{\rho_0}\nonumber\\
& &+B(\frac{\rho}{\rho_0})^{\sigma}(1-x\delta^2)-8x\tau\frac{B}{\sigma+1}\frac{\rho^{\sigma-1}}{\rho_0^\sigma}\delta\rho_{\tau'}\nonumber\\
& &+\frac{2C_{\tau,\tau}}{\rho_0}\int
d^3\,\vec{p^{'}}\frac{f_\tau(\vec{r},\vec{p^{'}})}{1+(\vec{p}-\vec{p^{'}})^2/\Lambda^2}\nonumber\\
& &+\frac{2C_{\tau,\tau'}}{\rho_0}\int
d^3\,\vec{p^{'}}\frac{f_{\tau'}(\vec{r},\vec{p^{'}})}{1+(\vec{p}-\vec{p^{'}})^2/\Lambda^2},
\label{buupotential}
\end{eqnarray}
where $\rho_0$ represents saturation density, and $\tau, \tau' = 1/2(-1/2)$ stand for neutron (proton). The $x$ parameter is incorporated to simulate various forms of symmetry energy predicted by different many-body theories without altering any symmetric nuclear matter property or symmetry energy at normal density. The mean field's short-range correlations are illustrated by parameters $A_u(x)$ = 33.037 - 125.34$x$ MeV, $A_l(x)$ = -166.963 + 125.34$x$ MeV, B = 141.96 MeV, $C_{\tau,\tau}$ = 18.177 MeV, $C_{\tau,\tau'}$ = -178.365 MeV, $\sigma = 1.265$, and $\Lambda = 630.24$ MeV/c \cite{yongsrc}. These values correspond to the empirical saturation density $\rho_{0}$ = 0.16 fm$^{-3}$, binding energy $E_{0}$ = -16 MeV, incompressibility $\kappa_{0}$ = 230 MeV, the isoscalar effective mass $m_{s}^{*} = 0.7 m$, the single-particle potential $U^{0}_{\infty}$ = 75 MeV at infinitely large nucleon momentum at saturation density in symmetric nuclear matter, and the symmetry energy $E_{\rm sym}(\rho_0) = 34.57$ MeV \cite{yongsrc}. These settings allow for an accurate reproduction of the experimental Hama potential at saturation density \cite{hama90}.
Hadron potentials for nucleons, resonances, hyperons and their antiparticles are deployed using the test-particle method \cite{yongrcas2022}. The kaon potential is derived from Ref.~\cite{ligq97}, and the pion potential is disregarded at relatively high energies \cite{pionp15}. For strange baryons, we employ the quark counting rule, which posits that these strange baryons interact with other baryons solely through their non-strange constituents \cite{mos74,chung2001}. Experimental data determine the free elastic proton-proton cross section, represented as $\sigma_{pp}$, and the neutron-proton cross section, represented as $\sigma_{np}$. The free elastic neutron-neutron cross section, represented as $\sigma_{nn}$, is presumed to be equivalent to the $\sigma_{pp}$ at a comparable center of mass energy. Furthermore, it is assumed that all other baryon-baryon free elastic cross sections are equal to the nucleon-nucleon elastic cross section at the same center of mass energy. An experimental energy-dependent nucleon-nucleon inelastic total cross-section is employed at lower energies \cite{nninel87}. The isospin-dependent baryon-baryon ($BB$) elastic and inelastic scattering cross sections in medium $\sigma_{BB}^{medium}$ are reduced compared to their free-space value $\sigma _{BB}^{free}$ by a factor of
\begin{eqnarray}
R^{medium}_{BB}(\rho,\delta,\vec{p})&\equiv& \sigma
_{BB}^{medium}/\sigma
_{BB}^{free}\nonumber\\
&=&(\mu _{BB}^{\ast }/\mu _{BB})^{2},
\end{eqnarray}
where $\mu _{BB}$ and $\mu _{BB}^{\ast }$ denote the reduced masses of the colliding baryon pairs in free space and medium, respectively \cite{yongsrc}. The updated AMPT-HC and AMPT-SM modes have been successfully applied to heavy-ion collisions at $\sqrt{s_{NN}}$ = 3 to 9 GeV range \cite{yong2023,wuzm2023}.


\begin{figure}[t]
\centering
\includegraphics[width=0.45\textwidth]{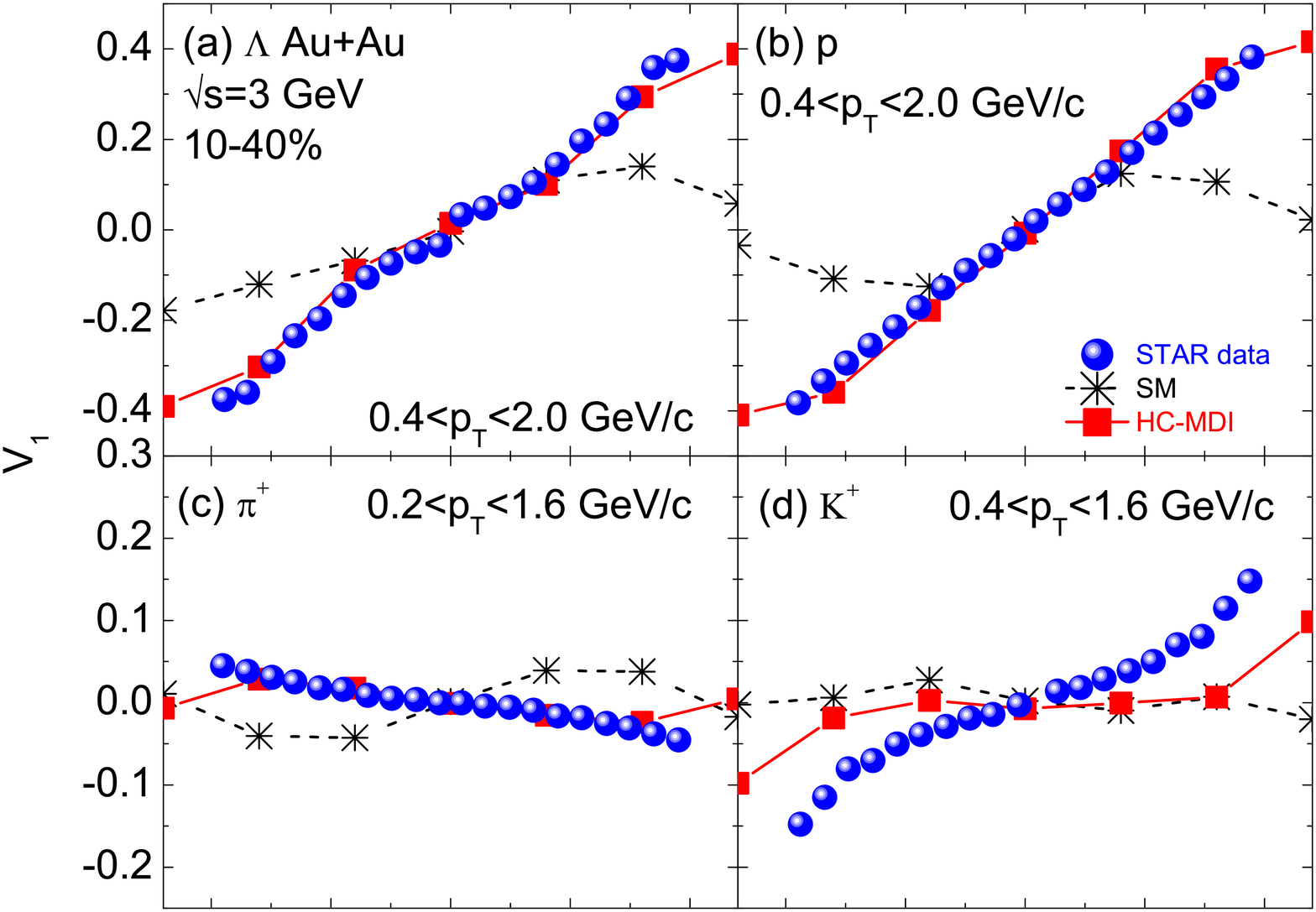}
\includegraphics[width=0.45\textwidth]{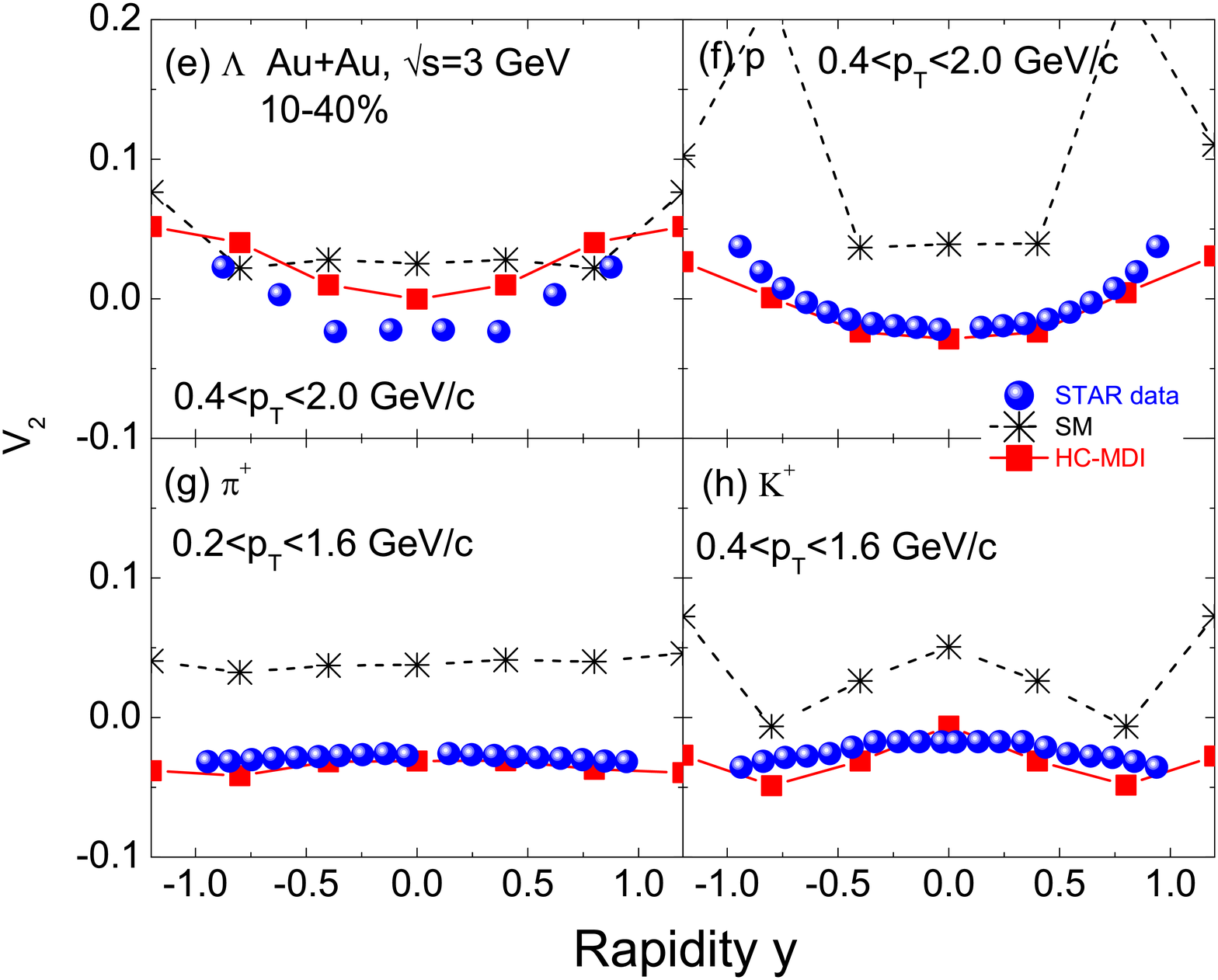}
\caption{Directed (top) and elliptic (bottom) flows of $\Lambda$, proton, $\pi^{+}$, and $K^{+}$ in 10-40\% centrality Au+Au collisions at $\sqrt{s_{NN}}$ = 3 GeV computed using AMPT-HC and AMPT-SM mode, respectively. Data are obtained from Ref.~\cite{mf2022}. Data at positive rapidities are reflections of the data in the negative region.} \label{flow3}
\end{figure}
To extract the underlying dynamic information from heavy-ion collisions, the particle differential distribution is often expressed as a Fourier series \cite{flow6,flow7,flow8}.
The directed flow, denoted as $v_{1}$, and the elliptic flow, represented by $v_{2}$, can be respectively expressed as:
$$
   v_{1}=\langle cos(\phi) \rangle=\langle \frac{p_{x}}{p_{T}} \rangle,
   v_{2}=\langle cos(2\phi) \rangle=\langle \frac{p_{x}^{2}-p_{y}^{2}}{p_{T}^{2}} \rangle,
$$
where $p_{T} = \sqrt{p_{x}^{2}+p_{y}^{2}}$ and $\phi$ represent the particle's transverse momentum and azimuthal angle, respectively. The directed flow ($v_{1}$) and elliptic flow ($v_{2}$) are commonly employed to investigate the characteristics of matter produced in high-energy nuclear collisions \cite{flow9,flow10,flow11}.

Figure~\ref{flow3} presents the rapidity distributions of $v_{1}$ and $v_{2}$ for $\Lambda$, proton, $\pi^{+}$, and $K^{+}$ in the 10-40\% centrality Au+Au collisions at $\sqrt{s_{NN}}$ = 3 GeV, using hadron transport (AMPT-HC) and quark transport (AMPT-SM), respectively. The hadron transport calculations closely match the STAR data for $v_{1}$ and $v_{2}$ of various hadrons, except for $v_{1}$ of $K^{+}$ and $v_{2}$ of $\Lambda$. The quark transport model, on the other hand, does not replicate the STAR data satisfactorily. However, the close fit of the AMPT-HC model's results to the STAR data at $\sqrt{s_{NN}}$ = 3 GeV validates the model's reliability in this energy range. Strikingly, other frequently employed models fail to produce satisfactory results \cite{qcd2022,eos2023}.
Figure~\ref{flow3} effectively highlights the generation of dense \emph{hadronic} matter in Au+Au collisions at $\sqrt{s_{NN}}$ = 3 GeV, in agreement with studies featured in Refs. \cite{mf2022,wuzm2023}.

\begin{figure}[t]
\centering
\includegraphics[width=0.45\textwidth]{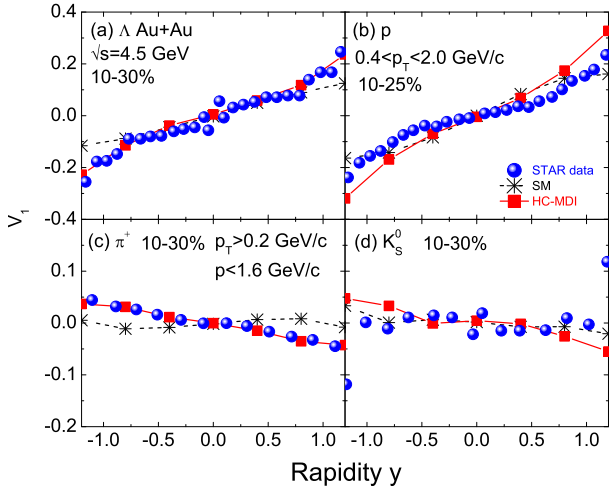}
\caption{Directed flows of $\Lambda$, proton, $\pi^{+}$, and $K^{0}_{S}$ in midcentral Au+Au collisions at $\sqrt{s_{NN}}$ = 4.5 GeV computed using both the AMPT-HC and AMPT-SM modes, respectively. The data have been sourced from Ref.~\cite{qk2021}.} \label{flow45}
\end{figure}
In Ref.~\cite{qk2021}, the STAR Collaboration puts forth the possibility of quark matter formation in Au+Au collisions at $\sqrt{s_{NN}}$ = 4.5 GeV through the analysis of pion and proton elliptic flows. They further suggest that a more conclusive determination can be achieved with the comprehensive statistics collected from the second phase of the Beam Energy Scan (BES-II). Figure~\ref{flow45} depicts the $v_{1}$ of $\Lambda$, proton, $\pi^{+}$, and $K^{0}_{S}$ in midcentral Au+Au collisions at $\sqrt{s_{NN}}$ = 4.5 GeV, as calculated through the hadron transport model AMPT-HC and the quark transport model AMPT-SM, respectively. Interestingly, aside from pion, the results from both transport models align closely and correspond well with the STAR data. With respect to the $v_{1}$ of $\pi^{+}$, the results from the hadron transport model match the data more aptly. Conversely, for $K^{0}_{S}$'s $v_{1}$, the quark transport model provides a better fit for the data. This isn't wholly unexpected considering that pions typically probe lower density compressed matter owing to prominent medium effects \cite{pion19}, while kaons generally examine the characteristics of denser matter at peak compression due to fewer interactions with neighboring nucleons \cite{cas2021}. Hence, it isn't surprising to see that the $v_{1}$ of a pion, simulated from the hadron transport model AMPT-HC, aligns well with collected data while the $v_{1}$ of a kaon, derived through the quark transport model AMPT-SM, also fits the data accurately. This suggests that both probes examine different density regions of nuclear matter created in heavy-ion collisions. Figure~\ref{flow45} illustrates the potential for the partial formation of quark matter in midcentral Au+Au collisions at $\sqrt{s_{NN}}$ = 4.5 GeV.

\begin{figure}[t!]
\centering
\includegraphics[width=0.5\textwidth]{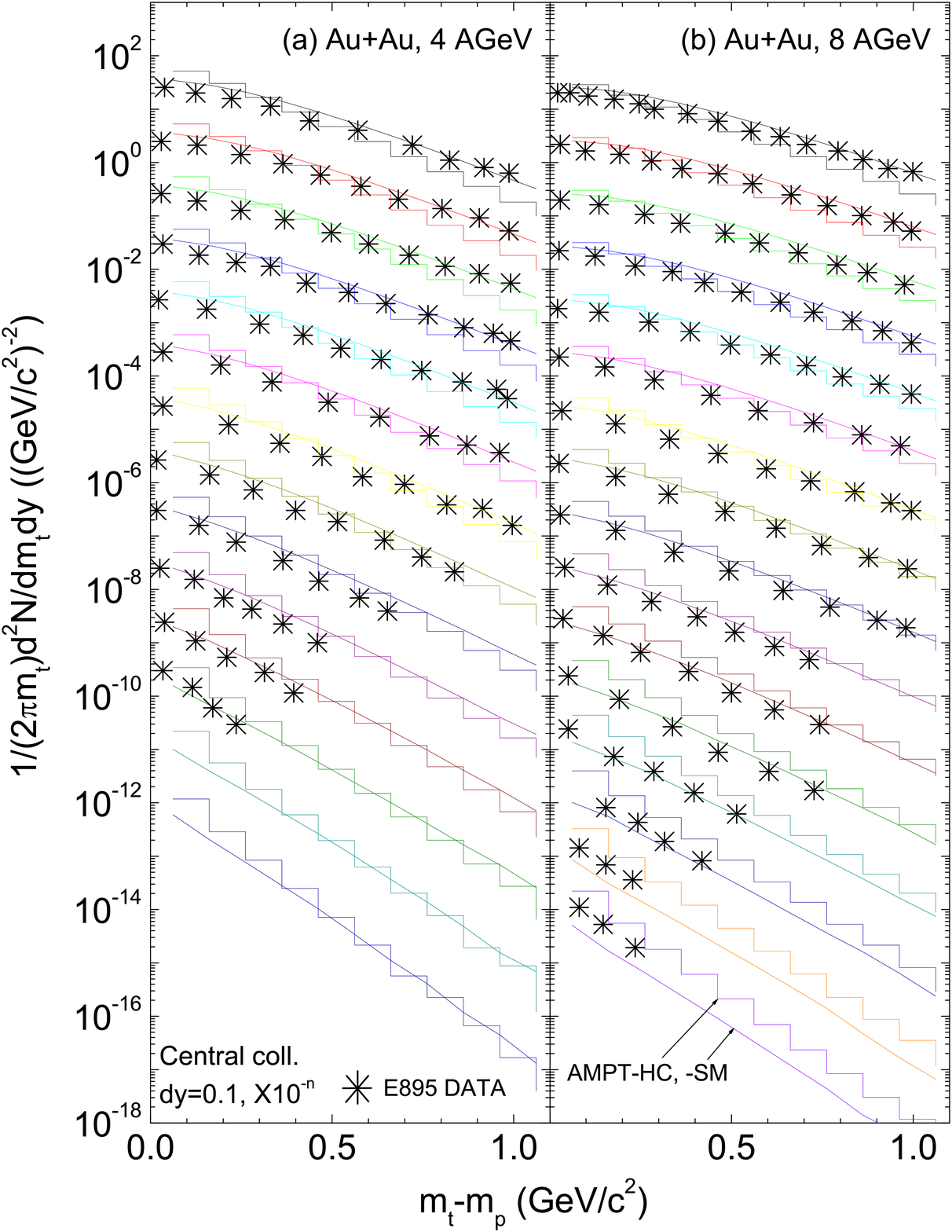}
\caption{Invariant yield per event for protons in central Au+Au collisions at 4 and 8 AGeV, with $\sqrt{s_{NN}}$ = 3.32 and 4.3 GeV respectively, is presented as a function of $m_{t}$-$m_{p}$ with both AMPT-HC (staircase line) and AMPT-SM (straight line). The midrapidity values are shown without scaling, while the forward/backward rapidity $|y|$ slices of 0.1 unit are scaled down successively by factors of 10. Data are taken from Ref.~\cite{mtdis}.} \label{mtp}
\end{figure}
\begin{figure}[t!]
\centering
\includegraphics[width=0.5\textwidth]{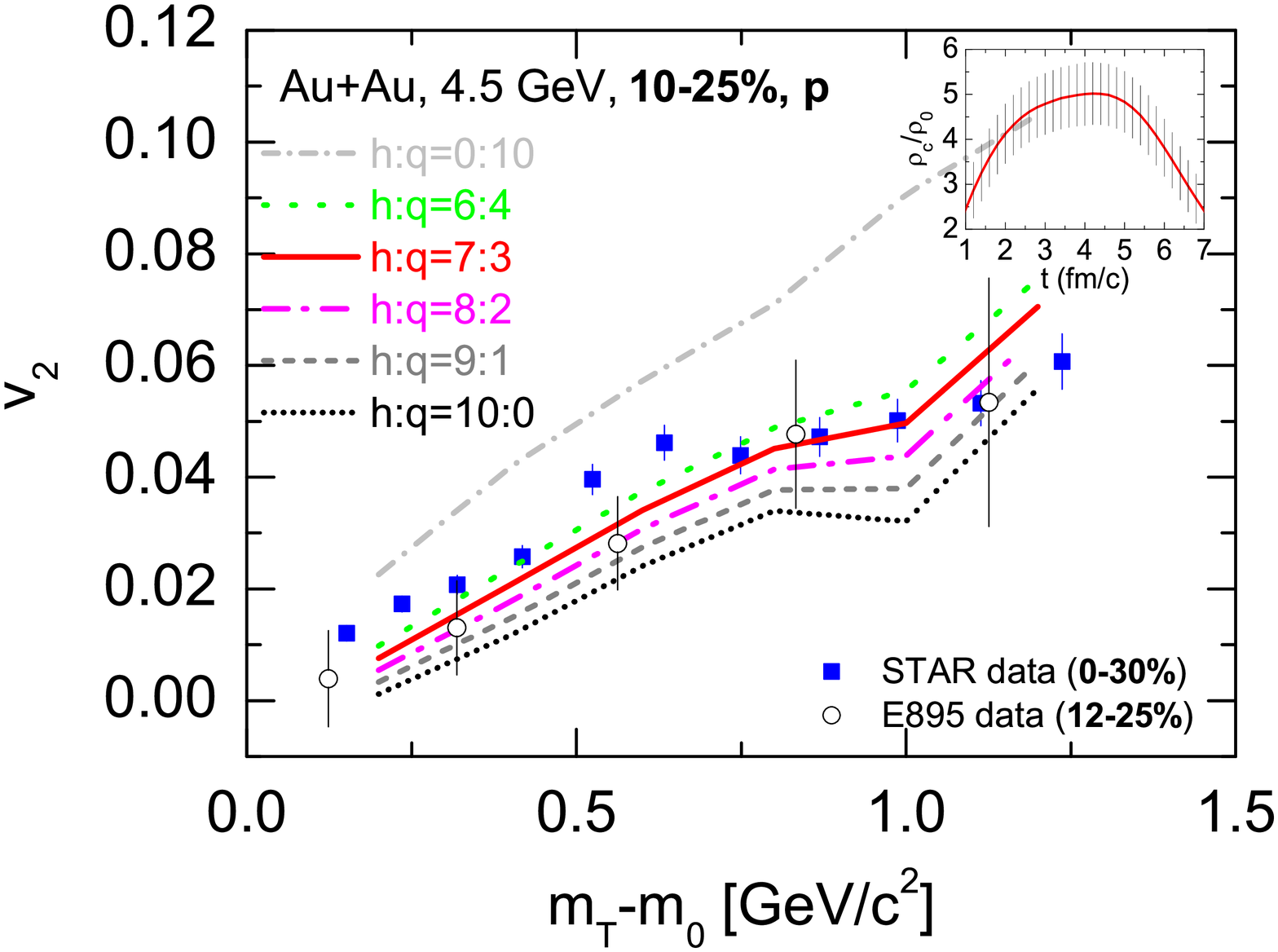}
\caption{Proton elliptic flows in midcentral Au+Au collisions at $\sqrt{s_{NN}}$ = 4.5 GeV, as simulated by AMPT-HC and AMPT-SM model-mixed events in different proportions (i.e., 0:10, 6:4, 7:3, 8:2, 9:1, and 10:0, respectively), are represented in this work. The inset displays the evolution of the compressed baryon density in the central cell, along with the synchronous representation of the standard deviation. Data regarding the elliptic flow of protons were sourced from Ref.~\cite{qk2021,e895p}.} \label{flowp}
\end{figure}
To clearly demonstrate the formation of quark matter in Au+Au collisions at $\sqrt{s_{NN}}$ = 4.5 GeV, a study is performed on proton elliptic flows by mixing events in varying proportions, as determined by the hadron and quark transports respectively. To validate the accuracy of the model employed in identifying a potential hadron-quark phase transition, additional checks and verifications are essential. A variety of tests should be carried out to verify the consistency and validity of the model predictions, ensuring their reliability. This increases our assurance in the model's competence to effectively identify the existence or lack thereof of a phase transition within the studied energy range. Figure~\ref{mtp} juxtaposes the invariant yield per event for protons in central Au+Au collisions within our energy range of interest. It is clear that both AMPT-HC and AMPT-SM can provide an apt representation of the data. However, the current comparison does not provide unambiguous evidence of a hadron-quark phase transition within these energy ranges. Figure~\ref{flowp} presents the proton elliptic flow results as a function of transverse mass. In this research, a combination of various event methods was utilized. The proportion of mixed events, derived from hadron transport and quark transport respectively, was regulated in order to determine the final proton elliptic flow, also as a function of transverse mass. It was observed that hadronic matter predominates, but approximately a quarter of the matter generated in Au+Au interactions at $\sqrt{s_{NN}}$ = 4.5 GeV is quark matter. The pion elliptic flow was not examined in this instance due to potential insufficient statistics, as suggested in Ref.~\cite{qk2021}. From Figure~\ref{flowp}, it can be inferred that quark matter is formed prior to or at $\sqrt{s_{NN}}$ = 4.5 GeV during Au+Au interactions. In simpler terms, the hadron-quark phase transition occurs before or when $\sqrt{s_{NN}}$ = 4.5 GeV in Au+Au collisions. The inset indicates the maximum compression encountered during the Au+Au interactions at $\sqrt{s_{NN}}$ = 4.5 GeV. At about $(5\pm 0.7)\rho_{0}$, equivalent to 5 times nuclear compression, the hadron-quark phase transition takes place. The corresponding temperature and baryon chemical potential are approximately 112 MeV and 586 MeV, respectively \cite{chemt06}. Therefore, analyzing proton elliptic flow in Au+Au collisions at $\sqrt{s_{NN}}$ = 4.5 GeV reveals that the hadron-quark phase transition partially happens at around a baryon density of $\rho\sim (5\pm 0.7)\rho_{0}$, i.e., temperature $T\sim$ 112 MeV and baryon chemical potential $\mu_{B}\sim$ 586 MeV, as depicted in the nuclear matter phase diagram in Figure~\ref{ptdia}. This establishes a phase transition boundary.

The current constrained phase transition boundary appears to align closely with the findings derived from the investigation of double strangeness production \cite{yong2023}. The triggering beam energy for the onset of the hadron-quark phase transition differs mainly due to the fact that protons and doubly strange baryons explore distinct regions of compression density \cite{yong2023}.

%
In summary, the exploration of hot and dense nuclear matter's phase diagram is a primary objective of relativistic heavy-ion collisions. The newly updated isospin and momentum-dependent hadron transport simultaneously reproduce the RHIC-STAR experimental data for the directed and elliptic flows of mesons and baryons in midcentral Au+Au collisions at $\sqrt{s_{NN}}$ = 3 GeV. Conversely, the proton elliptic flow data in midcentral Au+Au collisions at $\sqrt{s_{NN}}$ = 4.5 GeV can only be replicated by phase-mixing transport. Further research indicates that the hadron-quark phase transition may take place in Au+Au collisions at $\sqrt{s_{NN}}$ = 4.5 GeV, i.e., at roughly five times nuclear compression, corresponding to a temperature $T\sim$ 112 MeV and a baryon chemical potential $\mu_{B}\sim$ 586 MeV. The findings derived here may have important implications for the study of primordial black holes created in the early universe, neutron star structures, and the dynamical evolution of neutron star mergers. Furthermore, these results provide new insight into non-perturbative QCD in high baryon density regions.

%
This work is supported by the National Natural Science Foundation of China under Grant Nos. 12275322, 12335008 and the Strategic Priority Research Program of Chinese Academy of Sciences with Grant No. XDB34030000.

\end{document}